\documentclass[12pt,a4paper]{article}                                                        
\usepackage[cp1251]{inputenc}
\usepackage{indentfirst}
\usepackage{misccorr}
\usepackage{graphicx}
\usepackage{stackrel,amsmath}
\usepackage{authblk}
\usepackage{graphicx}

\begin{document}

\begin{titlepage}
\author{Junussov I.A.}
\affil{Institute of Information and\\ Computational Technologies, Almaty}

\title{Note on distance matrix hashing\thanks{This research is conducted 
within the framework of the program-targeted funding project num. BR05236839 
“Development of information technologies and systems for stimulation of 
personality's sustainable development as one of the bases of development of 
digital Kazakhstan”}}
\end{titlepage}

\maketitle                     
\begin{abstract}
Hashing algorithm of dynamical set of distances is described. Proposed hashing function is residual. 
Data structure which implementation accelerates computations is presented.
\end{abstract}

\section{Introduction}
Unweighted Pair Group Method with Centroid distance minimization~\cite{Sneath73}
or UPGMC is one of existing nonparametric clusterization algorithm. It starts with $n$ points 
(interpreted as clusters) in some coordinate (Euclidian) space, and on each of $n-1$ 
steps it merges closest pair of clusters into cluster which have coordinates of mass 
center of all points belonging to the pair. Details may be found in~\cite{Sneath73}. 

Simple program implementation of this algorithm recomputes distance matrix (elements 
above or under main diagonal) which shrinks on each step, this approach requires 
$O(n^3)$ distance computations, more precisely required number is tetrahedral 
 $T_{n-1}={n+1 \choose 3}$. 

Other possible way is to update set of actual distances without repeating already 
made calculations.
It requires computing of ${n \choose 2}$ 
initial distances and $n-2+n-3+\ldots+1={n-1 \choose 2}$ distances between merging clusters.
This sum ups to $(n-1)^2,$ i.e. $O(n^2)$ distance computations required.

Problem is to design data structure such that operations of distances updating and deletion 
would take reasonable time. This paper describes hashing (partition)~\cite{GrahamKnuthPatashnik}  
of dynamic set of distances instead of using matrix data type.  

\section{Description of data structure}
Let us enumerate points by consecutive numbers 
$\mathrm{id}_1,\mathrm{id}_2,\ldots,\mathrm{id}_k, k<\infty.$
 
Each element of dynamic set of distances is triple 
$(\mathrm{id}_m,\mathrm{id}_s,d_{ms})$ where $d_{ms}$ is 
distance between points indexed by $\mathrm{id}_m,\mathrm{id}_s.$ 
It is assumed that in any triple first element is less than second 
element. Dynamic set of distances is implemented as list $L$ of fixed
length $l.$ Each element of $L$ is dynamic list (slot) 
$S_j, j=0,\ldots,l-1.$ Thus, $L$ is:
$$
S_0 \to S_1 \to \ldots \to S_{l-1}.
$$

Triple $(\mathrm{id}_m,\mathrm{id}_s,d_{ms})$ belongs to slot $S_j$ if:  
\begin{equation}
\label{mod_eq}
j=(\mathrm{id}_m + \mathrm{id}_s) \mod l. 
\end{equation}
This is one of possible hashing functions on set containing ordered pairs of indexes.
In other words pair $(\mathrm{id}_m,\mathrm{id}_s)$ uniquely defines $j$ which is index
of slot $S_j \ni (\mathrm{id}_m,\mathrm{id}_s,d_{ms}).$
We use this partition of triples for acceleration of look-up, insertion and deletion in $L$.

Consider slot $S_j$ for some $j=0,1,\ldots,l-1$. Let it consist of following triples:
$$(\mathrm{id}_{x_1},\mathrm{id}_{y_1},d_{x_1 y_1}) \to (\mathrm{id}_{x_2},\mathrm{id}_{y_2},d_{x_2 y_2}) \to ... \to (\mathrm{id}_{x_p},\mathrm{id}_{y_p},d_{x_p y_p}).$$

Program implementation which creates and updates $S_j$ such that first elements of triples are sorted
in following way
\begin{equation}
\label{sort_ineq}
\mathrm{id}_{x_1} \leq \mathrm{id}_{x_2} \leq \ldots \leq \mathrm{id}_{x_p}
\end{equation}
allows to use binary search within slot. For example, if triple $(\mathrm{id}_m,\mathrm{id}_s,d_{ms})$ should be deleted:

\begin{enumerate}
\item find slot index by~\eqref{mod_eq}  
\item within found slot allocate first and last occurences of triples with first element equal to $\mathrm{id}_m$ by binary search
\item sequential search of $(\mathrm{id}_m,\mathrm{id}_s,d_{ms})$ within allocated sublist
\end{enumerate}

Adding of element is similar and should satisfy conditions~\eqref{mod_eq},~\eqref{sort_ineq}. It can be inserted at a position of first occurence from step 2 if found. 

If creation and updating of list additionally keeps nondecreasing sorting with respect to second component
for all triples with same first component within any slot then it allows to use binary search on step 3.
Theoretically, it can improve performance for large $n.$

\section{Conclusions}

Multiple executions of UPGMC with described hashing show considerable decrease of overall runtime compared to simple 
implementation mentioned in introduction. Overall runtime depends on number of slots $l.$ Number $l$ depends on $n$
and computational architecture. Parallelization of hashing algorithm is possible.

\end{document}